\documentclass[aps,twocolumn,superscriptaddress]{revtex4}%
\usepackage{amsfonts}
\usepackage{amsmath}
\usepackage{amssymb}
\usepackage{color}
\usepackage{graphicx}
\newcommand{\red}[1]{\textcolor{red}{#1}}

\begin{document}

\title{Spatial Kramers-Kronig relation and controlled unidirectional reflection in cold atoms}
\author{Yan Zhang}
\affiliation{School of Physics, Northeast Normal University, Changchun 130024, China}
\author{Jin-Hui Wu}
\email{jhwu@nenu.edu.cn}
\affiliation{School of Physics, Northeast Normal University, Changchun 130024, China}
\affiliation{State Key Laboratory of Quantum Optics and Quantum Optics Devices, Shanxi University, Taiyuan 030006, China}
\author{M. Artoni}
\affiliation{Department of Engineering and Information Technology and Istituto Nazionale di Ottica (INO-CNR), Brescia University, 25133 Brescia, Italy}
\author{G. C. La Rocca}
\affiliation{Scuola Normale Superiore and CNISM, 56126 Pisa, Italy}
\date{\today }

\begin{abstract}
We propose a model for realizing frequency-dependent spatial
variations of the probe susceptibility in a cold atomic
sample. It is found that the usual Kramers-Kronig (KK) relation
between real and imaginary parts of the probe susceptibility in the
frequency domain can be mapped into the space domain as a far
detuned control field of intensity linearly varied in space is
used. This non-Hermitian medium exhibits then a unidirectional
reflectionless frequency band for probe photons incident from either
the left or the right sample end. It is of special interest that we
can tune the frequency band as well as choose the direction
corresponding to the vanishing reflectivity by changing,
respectively, the control field intensity and frequency. The nonzero
reflectivity from the other direction is typically small for
realistic atomic densities, but can be largely enhanced by
incorporating the Bragg scattering into the spatial KK relation
so as to achieve a high reflectivity contrast.
\end{abstract}

\maketitle

\section{Introduction}

Asymmetric-reflection and unidirectional reflectionless control of
the flow of photons, a key technique for realizing photonic quantum
manipulation and communication, has attracted intense research
efforts because of its imminent applications in developing novel
photonic circuits and
devices~\cite{NPBGMV1,NPBGMV2,NPBGPT1,NPBGPT2,NPBGKK1,KKLoh1,NPBGKKExp1,NPBGKK2}.
The reflection control of light signals is usually reciprocal and
static (\textit{i.e.}, determined by growth design) as achieved,
\textit{e.g.}, via fixed band gaps of photonic crystals possessing
certain periodic structures of the \textit{real} refractive
index~\cite{PC1,PC2}. A tunable photonic band gap has been proved to
be viable by establishing controlled periodic structures of the
\textit{complex} susceptibility in the regime of electromagnetically
induced transparency (EIT)~\cite{EIT1,EIT2}, with standing-wave
coupling fields to dress homogeneous atomic
clouds~\cite{EPBG1,EPBG2,EPBG3,EPBG4,EPBG5} or traveling-wave
coupling fields to dress periodic atomic
lattices~\cite{ALPBG1,ALPBG2,ALPBG3,ALPBG4,ALPBG5,ALPBG6,ALPBGEIT1,ALPBGEIT2,ALPBGEIT3,ALPBGEIT4,ALPBGEIT5,ALPBGEIT6}.
Generally speaking, it is hard to achieve asymmetric light transport
in the familiar linear optical
processes~\cite{Nonreci,NonreciAPP1,NonreciAPP2}, though significant
progress has been made in the recent years by considering moving
atomic lattices~\cite{NPBGMV1,NPBGMV2,ALPBGEIT5} and fabricating
materials of parity-time (PT) symmetry or
asymmetry~\cite{NPBGPT1,NPBGPT2,NPBGPT3,NPBGPT4}. Experimental
implementations of these schemes, however, are rather challenging
due to the needs of complicated atom-light coupling configurations,
precise spatial field arrangement, and balanced gain and loss in a
single period.

Alternatively, asymmetric and unidirectional reflection can be
realized in an inhomogeneous continuous medium as the usual
Kramers-Kronig (KK) relation, satisfied by its complex optical
response function, is mapped from the frequency domain to the space
domain~\cite{NPBGKK1,KKLoh1,NPBGKKExp1,KKTheo0,KKExp1,NPBGKK2,KKLoh2,KKTheo1,KKTheo2,KKTheo3,KKLoh3,KKLoh4}.
This intriguing idea is first proposed by Horsley et
al.~\cite{NPBGKK1}, who found that a non-Hermitian medium would not
reflect radiation from one side for all incident angles if its
complex permittivity shows a space instead of a frequency
dependence. The spatial KK relation is also found to promise the
realization of omnidirectional perfect absorber~\cite{NPBGKKExp1}
and transmissionless media~\cite{KKTheo4}. When extended into
discrete lattices, complex potentials exhibiting the spatial KK
relation may further become invisible to support a bidirectional
reflectionless behavior~\cite{KKLoh3,KKLoh4}. These works not only
deepen our understanding of light propagation, but also provide a
new platform for realizing multi-functional optical elements,
especially those requiring perfect antireflection. In particular,
the original idea has been experimentally demonstrated via a
suitable design of different inhomogeneous
media~\cite{NPBGKKExp1,KKExp1}. Once again, these schemes have the
disadvantage of lacking dynamic tunability, being based on fixed
spatial structures of complex refractive index, susceptibility, or
permittivity.

Here we propose an efficient scheme for mapping the KK relation
of a probe susceptibility from the frequency domain into the space
domain in a cold atomic sample. The essence is to generate a
position-dependent ground level shift with a far detuned control
field of intensity linearly varied in space. Depending on the probe
frequency, the sample is found to exhibit the unbroken,
transitional, or broken regime in regard of the spatial KK
relation. The unbroken regime with a well-satisfied spatial KK
relation is of particular interest because it allows the
reflectionless manipulation of probe photons incident from one
sample end in a tunable spectral range. The transitional regime
with a partially-destroyed spatial KK relation is also of
interest because it may be explored to realize the
reflectionless manipulation from both sample ends, albeit at
different frequencies. More importantly, we can swap the
direction of vanishing reflectivity and that of nonzero reflectivity
by changing the sign of ground level shift, and enhance the nonzero
reflectivity while retaining the vanishing reflectivity by
increasing the magnitude of ground level shift. Last but not least,
the Bragg scattering in an atomic lattice can be incorporated into
the spatial KK relation to yield a high forward-backward
reflectivity contrast, corresponding to an enhanced nonzero
reflectivity and an invariant vanishing reflectivity.

\section{Model and Equations}

\begin{figure}[t]
\centering \includegraphics[width=\linewidth]{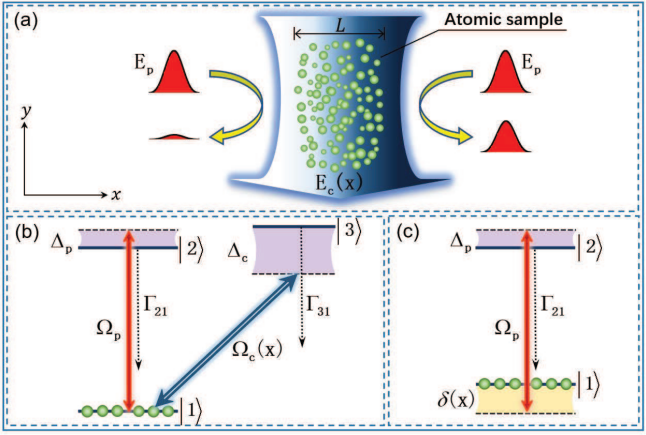} \caption{(a) A
cold atomic sample illuminated by a control beam $\mathbf{E}_{c}(x)$
along the $-y$ direction exhibits a strongly asymmetric reflection
for a probe beam $\mathbf{E}_{p}$ incident in the $\pm x$
directions. (b) A three-level atomic system driven by a probe field
of Rabi frequency $\Omega_{p}$ (detuning $\Delta_{p}$) and a control
field of Rabi frequency $\Omega_c(x)$ (detuning $\Delta_c$) into the
$V$ configuration. (c) A two-level atomic system with a dynamic
shift $\delta(x)$ of level $\vert 1\rangle $ upon the adiabatic
elimination of level $\vert 3\rangle $.} \label{fig:Fig1}
\end{figure}

We consider in Fig.~\ref{fig:Fig1}(a) a cold atomic sample extending
from $x=0$ to $x=L$, driven by a weak probe field of amplitude
(frequency) $\mathbf{E}_{p}$ ($\omega_{p}$) and a strong control
field of amplitude (frequency) $\mathbf{E}_{c}$ ($\omega_c$). The
control field is assumed to illuminate the sample along the $-y$
direction while the probe field can travel through the sample along
either $x$ or $-x$ direction. All atoms are driven into the
three-level $V$ configuration, as shown in Fig.~\ref{fig:Fig1}(b),
characterized by Rabi frequencies (detunings)
$\Omega_{p}=\mathbf{E}_{p}\cdot\mathbf{d}_{12}/2\hbar$ on transition
$\vert 1\rangle \leftrightarrow \vert 2\rangle $
($\Delta_{p}=\omega_{p}-\omega _{21}$) and
$\Omega_{c}=\mathbf{E}_{c}\cdot\mathbf{d}_{13}/2\hbar$ on transition
$\vert 1\rangle \leftrightarrow \vert 3\rangle $
($\Delta_{c}=\omega_{c}-\omega _{31}$), being
$\mathbf{d}_{\mu\nu}$ and $\omega _{\mu\nu}$ dipole moments and
resonant frequencies of relevant transitions. We have also used
$\Gamma _{31}$ and $\Gamma _{21}$ to describe the population decay
rates from levels $|3\rangle $ and $|2\rangle $ to level
$|1\rangle$, respectively. To be more concrete, levels
$|3\rangle $, $|2\rangle $, and $|1\rangle$ may refer to states
$\vert 5P_{3/2},F=3,m_{F}=3\rangle $, $\vert
5P_{1/2},F=1,m_{F}=1\rangle $, and $\vert
5S_{1/2},F=2,m_{F}=2\rangle $ of $^{87}$Rb atoms, respectively. This
choice ensures that ($i$) dipole moments $|\mathbf{d}_{12}|$ and
$|\mathbf{d}_{13}|$ take the largest possible values, which could
relax the requirement of a very dense atomic sample for achieving a
high reflectivity contrast; ($ii$) the control field doesn't couple
level $|1\rangle$ to a fourth level $|4\rangle$ even in the case of
a large $|\Delta_{c}|$, because no others except level $|3\rangle$
has $m_{F}=3$ on the $D_{2}$ line of $^{87}$Rb atoms. Most
importantly, we will assume that the control field is linearly
varied in intensity along the $x$ direction, \textit{e.g.}, by
a neutral density filter (NDF). In this case, $\vert \Omega_{c}\vert
^{2}$ should be replaced by $\vert \Omega_{c}(x)\vert ^{2}=x\vert
\Omega_{c0}\vert^{2}/L$ with $\Omega_{c0}$ denoting the maximal Rabi
frequency at $x=L$.

With the electric-dipole and rotating-wave approximations, working
in the weak probe limit, we can solve density matrix equations for
the three-level $V$ configuration to attain the steady-state probe
susceptibility
\begin{widetext}
\begin{equation}
\chi_{3}(\Delta_{p},x)=i\frac{N_{0}\vert \mathbf{d}_{12}\vert ^{2}}{\varepsilon _{0}\hbar }\frac{[(\gamma _{13}^{2}+\Delta _{c}^{2})
+\vert \Omega _{c}(x)\vert ^{2}] [\gamma_{23}-i(\Delta _{c}-\Delta _{p})]-\vert \Omega _{c}(x)\vert ^{2}(\gamma _{13}-i\Delta_{c})}
{[(\gamma_{13}^{2}+\Delta _{c}^{2}) +\vert\Omega _{c}(x)\vert ^{2}]
\{[\gamma _{23}-i(\Delta _{c}-\Delta _{p})](\gamma _{12}-i\Delta _{p})+\vert \Omega_{c}(x)\vert ^{2}\}},  \label{Eq1}
\end{equation}
\end{widetext}
where $N_{0}$ denotes the homogeneous atomic density while
$\gamma _{\mu\nu}$ is the coherence dephasing rate on
transition $|\mu\rangle \leftrightarrow |\nu\rangle $ with
$\gamma _{12}=\Gamma _{21}/2$, $\gamma _{13}=\Gamma _{31}/2$, and
$\gamma _{23}=(\Gamma_{31}+\Gamma _{21})/2$. Now we consider that
the control field is far detuned from transition $|1\rangle
\leftrightarrow |3\rangle $ by requiring $\Delta _{c}\gg
\Omega_{c0},\gamma_{13}$. In this case, level $\vert 3\rangle $ can
be adiabatically eliminated from the three-level $V$ configuration
to yield a two-level system [see Fig.~\ref{fig:Fig1}(c)], in which
level $ \vert 1\rangle $ suffers a position-dependent energy
shift $\delta (x)=\vert \Omega_{c}(x)\vert ^{2}/\Delta
_{c}=x\delta_{0}/L$ with $\delta _{0}=\vert \Omega_{c0}\vert
^{2}/\Delta _{c}$. It is worth noting that this level shift can
also be induced by a magnetic field of linearly varied magnitude in
space, which would however be less amenable to dynamic remote
control than an optical driving. Then the probe susceptibility can
be cast into a more compact form as given by
\begin{equation}
\chi_{2}(\Delta_{p},x) =i\frac{N_{0}\vert \mathbf{d}_{12}\vert ^{2}}{\varepsilon _{0}\hbar }\frac{1}{\gamma _{12}-i[\Delta_{p}+\delta (x)]},  \label{Eq2}
\end{equation}%
whose real ($\chi_{2}^{\prime}$) and imaginary
($\chi_{2}^{\prime\prime}$) parts govern the \textit{local}
dispersive and absorption properties around the probe resonance,
respectively. The possibility to control the effective detuning
$\Delta_{p}^{eff}(x)=\Delta_{p}+\delta(x)$ by changing the ground level shift $\delta(x)$ amounts to a direct mapping
from the spatially linear variation of the control intensity
to that of the probe detuning. The real and imaginary parts of
$\chi_{2}$ are well known to be related via the KK relation in the
frequency domain based on the causality principle and Cauchy's
theorem if we set $\delta(x)\equiv 0$~\cite{kk0}. Thus, for
appropriate probe detunings and sufficiently long samples such that
the spatial variation of the effective probe detuning induced
by $\delta(x)$ is fully developed, as shown by the red solid lines
in Figs.~\ref{Fig2}(a) and \ref{Fig2}(b), the KK relation holds in
the space domain and can be defined by the following Cauchy's
principal value integral
\begin{equation}
\chi_{2}^{\prime}(\Delta_{p},x)=\frac{1}{\pi}P\int_{0}^{L}\frac{\chi_{2}^{\prime\prime}(\Delta_{p},s)}{s-x}ds
\label{Eq3}
\end{equation}
over spatial coordinate $s$ along the $x$ direction.

A medium described by $\chi_{2}$ in Eq.~(\ref{Eq2}) is expected to
exhibit asymmetric light transport features, which can be examined
via the standard transfer-matrix method~\cite{TMM0,TMM1,TMM2} as
sketched below. First, the atomic sample is partitioned along the
$x$ direction into a large number ($J$) of slices labeled by
$j\in[1,J]$, which exhibit slightly different susceptibilities
$\bar{\chi}_{j}(\Delta_{p})=\chi_{2}(\Delta_{p},jl)$ but identical
length $l=L/J=10$ nm. Second, a $2\times 2$ unimodular transfer
matrix $M_{j}(\Delta_{p},l)$ characterized by $l$ and
$\bar{\chi}_{j}(\Delta_{p})$ can be established to describe
the propagation of an incident probe field of wavelength
$\lambda_{p}$ through the \red{$jth$} slice by 
\begin{equation} \left[
\begin{array}{c}
E_{p}^{+}(\Delta_{p},jl) \\
E_{p}^{-}(\Delta_{p},jl)%
\end{array}%
\right] =M_{j}(\Delta_{p},l) \left[
\begin{array}{c}
E_{p}^{+}(\Delta_{p},jl-l) \\
E_{p}^{-}(\Delta_{p},jl-l)%
\end{array}%
\right] ,  \label{Eq4}
\end{equation}
where $E_{p}^{+}$ and $E_{p}^{-}$ denote the forward and backward
components of the probe field, respectively. Third, the total
transfer matrix of the atomic sample turns out to be $M(\Delta
_{p},L)=M_{J}(\Delta _{p},l)\cdot \cdot \cdot
M_{j}(\Delta_{p},l)\cdot \cdot \cdot M_{1}(\Delta _{p},l)$ as a
multiplication of the individual transfer matrices of all atomic
slices. Finally, we can write the reflectivities ($R_{l}\ne R_{r}$)
and the transmissivities ($T=T_{l}=T_{r}$) as
\begin{eqnarray}
R_{l}(\Delta_{p},L) &=& \left\vert r_{l}(\Delta_{p},L)\right\vert ^{2}=\left\vert \frac{M_{(21)}(\Delta_{p},L)}{M_{(22)}(\Delta_{p},L)}\right\vert ^{2}, \notag  \\
R_{r}(\Delta_{p},L) &=& \left\vert r_{r}(\Delta_{p},L)\right\vert ^{2}=\left\vert \frac{M_{(12)}(\Delta_{p},L)}{M_{(22)}(\Delta_{p},L)}\right\vert ^{2},  \label{Eq5} \\
T(\Delta_{p},L) &=& \left\vert t(\Delta_{p},L)\right\vert ^{2}=\left\vert \frac{1}{M_{(22)}(\Delta_{p},L)}\right\vert^{2}  \notag
\end{eqnarray}
in terms of the four matrix elements of $M(\Delta _{p},L)$. Here the
subscripts `$l$' and `$r$' have been used to denote that the weak
probe field is incident from the left and right sides along the $x$
and $-x$ directions, respectively.

\section{Results and Discussion}

\begin{figure}[t]
\centering \includegraphics[width=\linewidth]{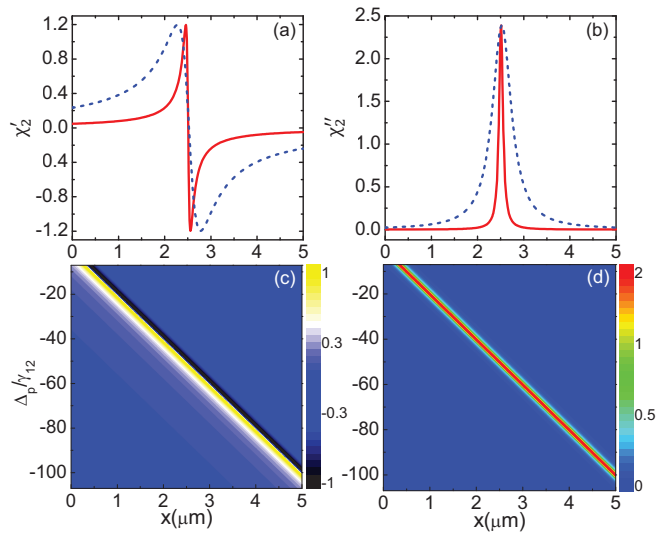} \caption{(a)
Real and (b) imaginary parts of susceptibility $\chi_{2}$ against
position $x$ with $\delta_{0}/\gamma_{12}$=100 and
$\Delta_p/\gamma_{12}=-50$ (red-solid); $\delta_{0}/\gamma_{12}$=20
and $\Delta_p/\gamma_{12}=-10$ (blue-dashed). (c) Real and (d)
imaginary parts of susceptibility $\chi_{2}$ against position $x$
and detuning $\Delta_p$ with $\delta_{0}/\gamma_{12}$=100. Other
parameters used in calculations are $\gamma_{12}=2.87$ MHz,
$d_{12}=1.79\times 10^{-29}$ C$\cdot$m, $N_{0}=2.0 \times 10^{13}$
cm$^{-3}$, $L=5.0$ $\mu$m, and $\lambda_{p}=795$ nm.} \label{Fig2}
\end{figure}

In this section, we show via numerical calculations how to
implement the spatial KK relation in a narrow spectral range by
tailoring the complex probe susceptibility, and how to implement the
unidirectional reflection of a high reflectivity contrast by
utilizing the spatial KK relation. All numerical calculations will
be done with realistic parameters corresponding to the three states
of $^{87}$Rb atoms chosen above, though our driving configuration
can also be realized, \textit{e.g.}, in other alkali metal atoms.

First, we plot in Figs.~\ref{Fig2}(a) and \ref{Fig2}(b),
respectively, the real and imaginary parts of $\chi_{2}$ against
position $x$ for two sets of parameters making the choice
$\Delta_{p}=-\delta_{0}/2$, which allows the effective probe
detuning $\Delta_{p}^{eff}(x)$ to be in the range of
$\{-\delta_{0}/2,\delta_{0}/2\}$. It is clear that
$\chi_{2}^{\prime}$ and $\chi_{2}^{\prime\prime}$ show an odd
profile and an even profile, respectively, centered at $z=L/2$ and
practically fully contained by the atomic sample, so they should
satisfy the spatial KK relation described by Eq.~(\ref{Eq3}). We
also can see that smaller (larger) values of
$\delta_{0}/\gamma_{12}$, a key dimensionless parameter in our
reduced two-level system, will result in broader (narrower) spatial
profiles of $\chi_{2}^{\prime}$ and $\chi_{2}^{\prime\prime}$ yet
without changing their peak amplitudes. To reveal the
frequency-dependent feature, we plot in Figs.~\ref{Fig2}(c) and
\ref{Fig2}(d), respectively, the real and imaginary parts of
$\chi_{2}$ against both position $x$ and detuning $\Delta_{p}$
instead. The profiles of $\chi_{2}^{\prime}$ and
$\chi_{2}^{\prime\prime}$ are found to move simultaneously toward
the left (right) sample end with the increasing (decreasing) of
$\Delta_{p}$. Accordingly, the spatial KK relation will be gradually
destroyed because Eq.~(\ref{Eq3}) becomes less and less satisfied.

In order to assess the extent to which the spatial KK relation is
satisfied for different probe detunings in a finite atomic sample,
here we propose the following integral
\begin{equation}
D_{kk}(\Delta_{p})=\frac{\int_{0}^{L}\left\{\chi_{2}^{\prime}(\Delta_{p},x) -\frac{1}{\pi}P\int_{0}^{L}\frac{\chi_{2}^{\prime\prime}(\Delta_{p},s)}
{s-x}ds\right\} dx}{\left\vert \int_{0}^{L}\chi_{2}^{\prime}(\Delta_{p},x)dx\right\vert } \label{Eq4}
\end{equation}
as a figure of merit for the spatial KK relation. In this
definition, $D_{kk}=0$ denotes the unbroken regime where the spatial
KK relation is fully satisfied while $D_{kk}=\pm1$ denote the broken
regime where the spatial KK relation is fully destroyed. According
to Eq.~(\ref{Eq2}) and Fig.~\ref{Fig2}, the spatial profiles of
$\chi_{2}^{\prime}$ and $\chi_{2}^{\prime\prime}$ are well contained
within our finite atomic sample only for a small range of
$\Delta_{p}$, so the validity of the spatial KK relation is expected
to increasingly deteriorate ($D_{kk}=0\to D_{kk}=\pm1$) as
$\Delta_{p}$ is gradually modulated out of this range. This is
different from all previous works on spatial KK
relations~\cite{NPBGKK1,KKLoh1,NPBGKKExp1,KKTheo0,KKExp1,NPBGKK2,KKLoh2,KKTheo1,KKTheo2,KKTheo3,KKLoh3,KKLoh4},
where the susceptibility or permittivity has been assumed to be
fixed by design, \textit{i.e.} not tunable.

\begin{figure}[t]
\centering \includegraphics[width=\linewidth]{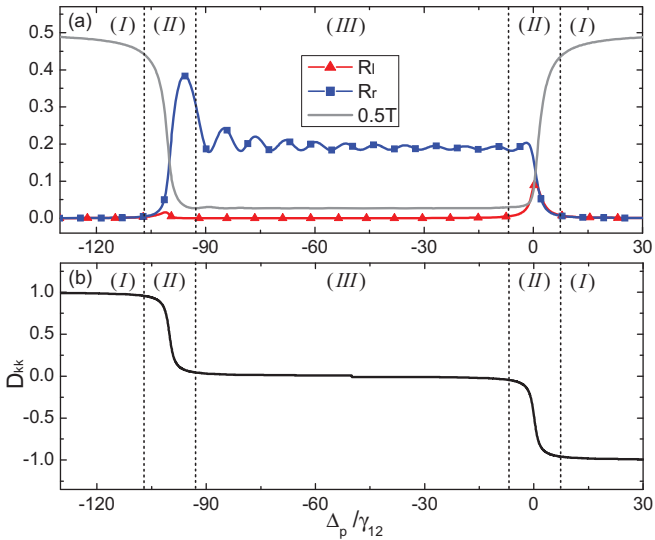} \caption{(a)
Reflectivity $R_{l}$, reflectivity $R_{r}$, and half transmissivity
$T$; (b) figure of merit $D_{kk}$ against detuning $\Delta _{p}$.
Other parameters used in calculations are the same as in
Fig.~\ref{Fig2}(a,b).} \label{Fig3}
\end{figure}

Then we plot in Fig.~\ref{Fig3}(a) the reflection and transmission
spectra for the parameters used in Figs.~\ref{Fig2}(a) and
\ref{Fig2}(b) based on Eq.~(\ref{Eq5}). It is easy to see that
these spectra can be divided into three regions: ($I$) where we have
$R_{l}=R_{r}\to 0$ and $T\to 1$; ($II$) where $R_{l}\ne R_{r}$ and
$T$ are sensitive to $\Delta_{p}$; ($III$) where $T\simeq 0.05$,
$R_{l}\to 0$, but $R_{r}$ oscillates around $0.2$. The generation
of three different regions can be understood by examining in
Fig.~\ref{Fig3}(b) the figure of merit $D_{kk}$ against probe
detuning $\Delta_{p}$, which clearly shows, as compared to
Fig.~\ref{Fig3}(a), that $D_{kk}$ governs the relations between
$R_{l}$, $R_{r}$, and $T$. The symmetric ($I$), asymmetric
($II$), and unidirectional ($III$) reflection regions correspond,
respectively, to the broken ($D_{kk}=\pm1$), transitional
($0<|D_{kk}|<1$), and unbroken ($D_{kk}=0$) regimes, and to the
cases when the absorption ($\chi_{2}^{\prime\prime}$) and dispersion
($\chi_{2}^{\prime}$) profiles move out of the sample, lie at the
sample boundaries, and are well contained by the sample. It is
worth noting that in region ($III$) a left (right) incident probe
beam is reflectionless (partially reflected) because it first sees
the negative (positive) peak of $\chi_{2}^{\prime}$~\cite{NPBGKK1},
and the resonant absorption ($\chi_{2}^{\prime\prime}$) is already
strong enough to yield $T \to 0.0$ for forward photons while the
dispersion profile ($\chi_{2}^{\prime}$) is not too sharp to yield
$R_{r}\to 1.0$ for backward photons. One way for further reducing
$T$ and simultaneously increasing $R_{r}$ is to produce enhanced
absorption profiles and sharper dispersion profiles in denser
atomic samples. The restricted range of densities of cold atoms
available in experiment, however, places a constraint on this
approach.

\begin{figure}[t]
\centering \includegraphics[width=\linewidth]{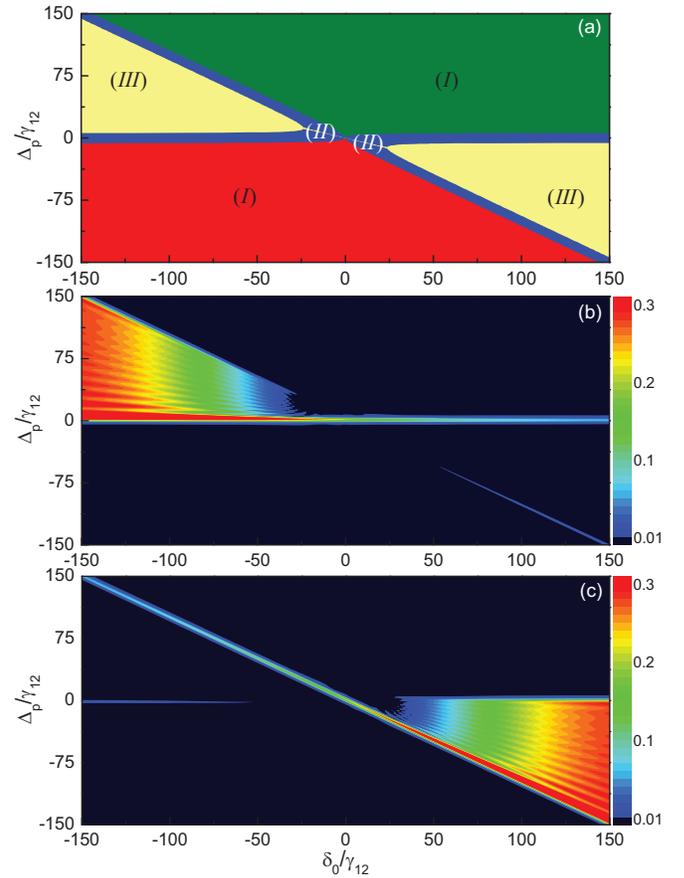} \caption{(a)
Figure of merit $D_{kk}$, (b) reflectivity $R_{l}$, and (c)
reflectivity $R_{r}$ (c) against shift $\delta_{0}$ and detuning
$\Delta _{p}$. Other parameters used in calculations are the same as
in Fig.~\ref{Fig2}(a,b).} \label{Fig4}
\end{figure}

Fig.~\ref{Fig4}(a) further shows the different regimes on a diagram
with $D_{kk}$ plotted against $\delta _{0}$ and $\Delta _{p}$, in which the green region ($D_{kk}=-1$) and the red region
($D_{kk}=1$) refer to the broken regime ($I$); the four narrow blue
regions ($0<|D_{kk}|<1$) refer to the transitional regime ($II$);
the two triangular yellow regions ($D_{kk}=0$) refer to the unbroken
regime ($III$). It is also clear that the widths of two yellow
regions depend critically on the magnitude of $\delta_{0}$; a
broken regime may be converted into an unbroken regime and vice
versa for a given $\Delta_{p}$ by changing the sign of $\delta_{0}$.
Accordingly, it is viable to enlarge or reduce the reflectionless
frequency band by varying the magnitude of $\delta_{0}$ and
convert the sample from left reflectionless to right reflectionless
or vice versa by changing the sign of $\delta_{0}$. This potentially dynamic controllability, a chief feature of our
proposal, is well demonstrated in Figs.~\ref{Fig4}(b) and
\ref{Fig4}(c) in terms of reflectivities $R_{l}$ and $R_{r}$.

\begin{figure}[t]
\centering \includegraphics[width=\linewidth]{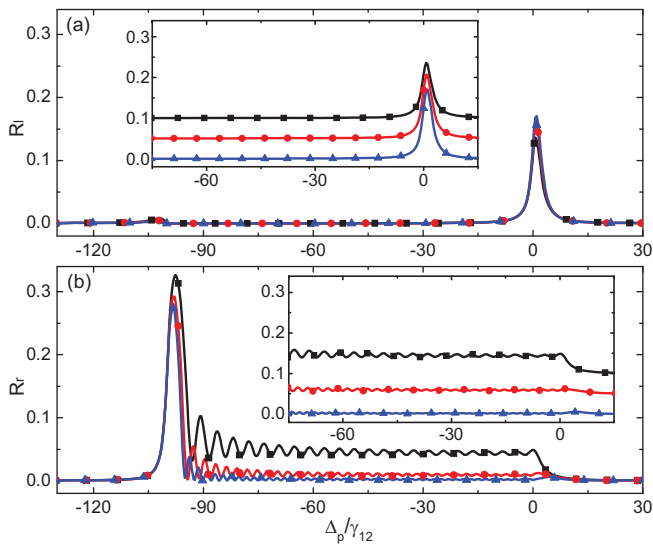}
\caption{Reflectivities (a) $R_{l}$ and (b) $R_{r}$ against detuning
$\Delta _{p}$ for $L=10$ $\mu$m (black-squares), $L=15$ $\mu$m
(red-circles), and $L=20$ $\mu$m (blue-triangles). Other parameters
used in calculations are the same as in Fig.~\ref{Fig2}(a,b).
Black (red) curves are shown with a vertical offset $0.1$
($0.05$) in both insets.} \label{Fig5}
\end{figure}

It is also interesting to examine what could happen for
reflectivities $R_{l}$ and $R_{r}$ when sample length $L$ is
multiplied while atomic density $N_{0}$ remains invariant. In this
case, we can see from $\delta(x)=x\delta_{0}/L$ that the linear
variation occurs in a much larger range while its magnitude
$\delta_{0}$ is unchanged. Then, as shown in Fig.~\ref{Fig5}, the
spatially wider/smoother dispersion ($\chi_{2}^{\prime}$) and
absorption ($\chi_{2}^{\prime\prime}$), of $L$-independent maxima
and minima, together result in a notable reduction of $R_{r}$ while
$R_{l}$ remains vanishing in the unbroken regime. The peak of
$R_{l}$ ($R_{r}$) accompanied by $R_{r}\to 0$ ($R_{l}\to 0$) in the
transitional regime has a $L$-independent position because it only
appears as the main profiles of $\chi_{2}^{\prime}(x)$ and
$\chi_{2}^{\prime\prime}(x)$ approach and even partially leave the
left (right) sample end. It is clearly not a result of the
spatial KK relation and allows two probe beams of different
frequencies to be simultaneously reflected or not when they are
incident upon the opposite sample ends. The damped oscillations of
$R_{r}$ against $\Delta_{p}$ in the unbroken regime can be
understood as a multiple interference effect due to the
discontinuities of the probe susceptibility at the right sample end
and at the resonant position inside the sample. It is clear that
stronger (weaker) oscillations occur at larger (smaller) values of
$|\Delta_{p}|$ because the resonant position of
$\chi_{2}^{\prime}(x)$ and $\chi_{2}^{\prime\prime}(x)$ is close to
(far from) the right sample end, yielding thus stronger (weaker)
discontinuities. The oscillation period can be roughly estimated as
$d\Delta_{p}\simeq \delta_{0}/L\cdot\lambda_{p}/2$ by considering
that the interval $d\Delta_{p}$ of two adjacent maxima corresponds
to a $2\pi$ phase shift ($\lambda_{p}/2$ spatial shift) gained by
the reflected photons ($\chi_{2}^{\prime}(x)$ and
$\chi_{2}^{\prime\prime}(x)$).

\begin{figure}[t]
\centering \includegraphics[width=\linewidth]{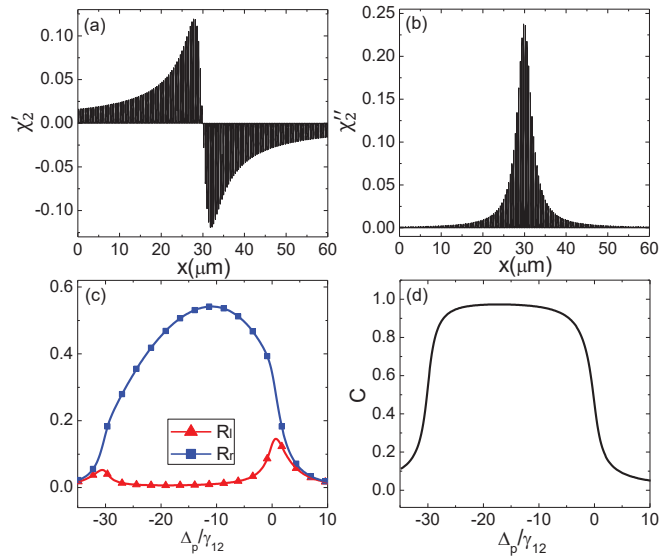} \caption{Spatial
profiles of (a) $\chi_{2}^{\prime}$ and (b)
$\chi_{2}^{\prime\prime}$ as well as corresponding spectra of (c)
reflectivities $R_{l,r}$ and (d) contrast $C$ for an atomic lattice
of density $N_{j}(x)=N_{0}e^{-(x-x_{j})^{2}/\delta x^{2}}$ in the
$jth$ trap of center $x_{j}=(j-1/2)a$, width $\delta x$, and
period $a$. Parameters are the same as in Fig.~\ref{Fig2}(b) except
$N_{0}=2.0 \times 10^{12}$ cm$^{-3}$, $L=60$ $\mu$m,
$\delta_{0}/\gamma_{12}=30$, $a=400$ nm, and $\delta x=a/6$.}
\label{Fig6}
\end{figure}

Finally, we note that an experiment may typically have a lower
atomic density and a larger sample length than in simulations
presented so far. To overcome this difficulty, we need to find an
alternative way to enhance the nonzero reflectivity in the unbroken
regime. This can be done by loading cold atoms into a 1D optical
lattice to create a spatially periodic density $N_{j}(x)$ as
described in the caption of Fig.~\ref{Fig6}, yielding thus Bragg
scattering incorporated into the spatial KK relation. As shown in
Fig.~\ref{Fig6}(a) and Fig.~\ref{Fig6}(b), both dispersion
$\chi_{2}^{\prime}(x)$ and absorption $\chi_{2}^{\prime\prime}(x)$
of one-order lower values now exhibit the comb-like spatial profiles
while satisfying to a less extent the spatial KK relation. In this
case, we can find from Fig.~\ref{Fig6}(c) that $R_{l}$ and $R_{r}$
are strongly asymmetric in a much smaller frequency range,
\textit{e.g.}, with $R_{r}$ exhibiting a maximal value up to $0.54$
while $R_{l}\lesssim 0.01$ for
$-25.5\lesssim\Delta_{p}/\gamma_{21}\lesssim -11.0$.
Fig.~\ref{Fig6}(d) further shows that the reflectivity contrast
$C=(R_{r}-R_{l})/(R_{r}+R_{l})$, an important figure of merit
on the asymmetric reflection, could be up to $0.97$ and is over
$0.90$ for $-27.0\lesssim\Delta_{p}/\gamma_{21}\lesssim -4.5$. It is
noticeable that the incorporation of Bragg scattering, typically
yielding symmetric reflectivities, has negligible effects on the
vanishing reflectivity but largely enhances the nonzero reflectivity
and the reflectivity contrast. That means, replacing a constant
density $N_{0}$ with a periodic density $N_{j}(x)$ does not hamper
the implementation of spatial KK relation, which is essential for
developing nonreciprocal optical devices requiring a high
reflectivity contrast.

It has been shown that atoms could be trapped and guided using
nanofabricated wires and surfaces to form atom chips~\cite{chip1}.
These chips provide a versatile experimental platform with cold
atoms and constitute the basis for wide and robust applications
ranging from atom optics to quantum optics. They have been used, for
instance, in diverse experiments involving quantum simulation,
metrology, and information processing~\cite{chip2,chip3,chip4}. We
then believe that our proposal is well poised to atom-chip
implementations in integrated optical devices.

\section{Conclusions}

In summary, we have investigated the spatial KK relation and
relevant reflection features in a short and dense sample of cold
$^{87}$Rb atoms. This nontrivial relation in regard of the probe
susceptibility is enabled by generating a position-dependent ground
level shift $\delta(x)$ with a far detuned control field of
intensity linearly varied along the $x$ direction. We find, in
particular, that the figure of merit $D_{kk}$ characterizing the
spatial KK relation may switch from the unbroken regime of
unidirectional reflection, via the transitional regime of asymmetric
reflection, to the broken regime of symmetric reflection, or vice
versa. This is attained by increasing the maximal level shift
$\delta_{0}$ from a negative value to a positive value or
considering an inverse process, depending on the sign of probe
detuning $\Delta_{p}$. A swapping between the nonzero reflectivity
and the vanishing reflectivity at opposite sample ends is also
viable by changing the sign of maximal level shift $\delta_{0}$. It
is of more interest that the nonzero reflectivity can be well
enhanced to result in a high reflectivity contrast for lower
densities and larger lengths in a cold atomic lattice, indicating
that Bragg scattering does not hamper or spoil the main effects of
spatial KK relation.

\section*{ACKNOWLEDGMENTS}

The work is supported by the National Natural Science Foundation of
China (No. 11534002, No. 11674049, and No. 11704064), the
Cooperative Program by the Italian Ministry of Foreign Affairs and
International Cooperation (No. PGR00960) and the National Natural
Science Foundation of China (No. 11861131001), the Jilin Scientific
and Technological Development Program (No. 20180520205JH), the
Program of State Key Laboratory of Quantum Optics and Quantum Optics
Devices (No. KF201807), and the Fundamental Research Funds for
Central Universities (2412019FZ045).


\begin{thebibliography}{99}
\bibitem{NPBGMV1} D.-W. Wang, H.-T. Zhou, M.-J. Guo, J. X. Zhang, J. Evers, and S.-Y. Zhu, Optical diode made from a moving photonic crystal, Phys. Rev. Lett. 110, 093901 (2013).

\bibitem{NPBGMV2} S. A. R. Horsley, J.-H. Wu, M. Artoni, and G. C. La Rocca, Optical nonreciprocity of cold atom bragg mirrors in motion, Phys. Rev. Lett. 110, 223602 (2013).

\bibitem{NPBGPT1} J.-H. Wu, M. Artoni, and G. C. La Rocca, Non-Hermitian degeneracies and unidirectional reflectionless atomic lattices, Phys. Rev. Lett. 113, 123004 (2014).

\bibitem{NPBGPT2} J.-H. Wu, M. Artoni, and G. C. La Rocca, Parity-time-antisymmetric atomic lattices without gain, Phys. Rev. A 91, 033811 (2015).

\bibitem{NPBGKK1} S. A. R. Horsley, M. Artoni, and G. C. La Rocca, Spatial Kramers-Kronig relations and the reflection of waves, Nat. Photonics 9, 436-439 (2015).

\bibitem{KKLoh1} S. Longhi, Bidirectional invisibility in Kramers-Kronig optical media, Opt. Lett. 41, 3727-3730 (2015).

\bibitem{NPBGKKExp1} W. Jiang, Y. Ma, J. Yuan, G. Yin, W. Wu, and S. He, Deformable broadband metamaterial absorbers engineered with an analytical spatial Kramers-Kronig permittivity profile, Laser Photonics Rev. 11, 1600253 (2017).

\bibitem{NPBGKK2} S. A. R. Horsley and S. Longhi, Spatiotemporal deformations of reflectionless potentials, Phys. Rev. A 96, 023841 (2017).

\bibitem{PC1} S. John, Strong localization of photons in certain disordered dielectric superlattices, Phys. Rev. Lett. 58, 2486-2489 (1987).

\bibitem{PC2} K. Sakoda, Optical properties of photonic crystals (Springer, Berlin, 2001).

\bibitem{EIT1} S. E. Harris, Electromagnetically induced transparency, Phys. Today 50(7), 36-42 (1997).

\bibitem{EIT2} M. Fleischhauer, A. Imamoglu, and J. P. Marangos, Electromagnetically induced transparency: Optics in coherent media, Rev. Mod. Phys. 77, 633-673 (2005).

\bibitem{EPBG1} A. Andr\'{e} and M. D. Lukin, Manipulating light pulses via dynamically controlled photonic band gap, Phys. Rev. Lett. 89, 143602 (2002).

\bibitem{EPBG2} X.-M. Su and B. S. Ham, Dynamic control of the photonic band gap using quantum coherence, Phys. Rev. A 71, 013821 (2005).

\bibitem{EPBG3} M. Artoni and G. C. La Rocca, Optically tunable photonic stop bands in homogeneous absorbing media, Phys. Rev. Lett. 96, 073905 (2006).

\bibitem{EPBG4} J. Wu, Y. Liu, D.-S. Ding, Z.-Y. Zhou, B.-S. Shi, and G. C. Guo, Light storage based on four-wave mixing and electromagnetically induced transparency in cold atoms, Phys. Rev. A 87, 013845 (2013).

\bibitem{EPBG5} Y. Zhang, Y.-M. Liu, X.-D. Tian, T.-Y. Zheng, and J.-H. Wu, Tunable high-order photonic band gaps of ultraviolet light in cold atoms, Phys. Rev. A 91, 013826 (2015).

\bibitem{ALPBG1} G. Birkl, M. Gatzke, I. H. Deutsch, S. L. Rolston, and W. D. Phillips, Bragg scattering from atoms in optical lattices, Phys. Rev. Lett. 75, 2823 (1995).

\bibitem{ALPBG2} I. H. Deutsch, R. J. C. Spreeuw, S. L. Rolston, and W. D. Phillips, Photonic band gaps in optical lattices, Phys. Rev. A 52, 1394-1410 (1995).

\bibitem{ALPBG3} P. M. Visser, G. Nienhuis, Band gaps and group velocity in optical lattices, Opt. Commun. 136, 470 (1997).

\bibitem{ALPBG4} S. Slama, C. von Cube, M. Kohler, C. Zimmermann, and Ph. W. Courteille, Multiple reflections and diffuse scattering in Bragg scattering at optical lattices, Phys. Rev. A 73, 023424 (2006).

\bibitem{ALPBG5} M. Antezza and Y. Castin, Fano-Hopfield model and photonic band gaps for an arbitrary atomic lattice, Phys. Rev. A 80, 013816 (2009).

\bibitem{ALPBG6} A. Schilke, C. Zimmermann, P. W. Courteille, and W. Guerin, Photonic band gaps in one-dimensionally ordered cold atomic vapors, Phys. Rev. Lett. 106, 223903 (2011).

\bibitem{ALPBGEIT1} D. Petrosyan, Tunable photonic band gaps with coherently driven atoms in optical lattices, Phys. Rev. A 76, 053823 (2007).

\bibitem{ALPBGEIT2} D. Yu, Photonic band structure of the three-dimensional $^{88}$Sr atomic lattice, Phys. Rev. A 84, 043833 (2011).

\bibitem{ALPBGEIT3} A. Schilke, C. Zimmermann, and W. Guerin, Photonic properties of one-dimensionally-ordered cold atomic vapors under conditions of electromagnetically induced transparency, Phys. Rev. A 86, 023809 (2012).

\bibitem{ALPBGEIT4} H. Yang, L. Yang, X.-C. Wang, C.-L. Cui, Y. Zhang, and J.-H. Wu, Dynamically controlled two-color photonic band gaps via balanced four-wave mixing in one-dimensional cold atomic lattices, Phys. Rev. A 88, 063832 (2013).

\bibitem{ALPBGEIT5} L. Yang, Y. Zhang, X.-B. Yan, Y. Sheng, C.-L. Cui, and J.-H. Wu, Dynamically induced two-color nonreciprocity in a tripod system of a moving atomic lattice, Phys. Rev. A 92, 053859 (2015).

\bibitem{ALPBGEIT6} Y. Zhang, Y.-M. Liu, T.-Y. Zheng, and J.-H. Wu, Light reflector, amplifier, and splitter based on gain-assisted photonic band gaps, Phys. Rev. A 94, 013836 (2016).

\bibitem{Nonreci} P. Yeh, Optical Waves in Layered Media (Wiley-Interscience, New York, 2005).

\bibitem{NonreciAPP1} L. Chang, X. Jiang, S. Hua, C. Yang, J. Wen, L. Jiang, G. Li, G. Wang, and Min Xiao, Parity-time symmetry and variable optical isolation in active-passive-coupled microresonators, Nat. Photon. 8, 524-529 (2014).

\bibitem{NonreciAPP2} B. Peng, S. K. $\mathrm{\ddot{O}}$zdemir, F. Lei, F. Monifi, M. Gianfreda, G. L. Long, S. Fan, F. Nori, C. M. Bender, and L. Yang, Parity-time-symmetric whispering-gallery microcavities, Nat. Phys. 10, 394-398 (2014).

\bibitem{NPBGPT3} Z. Lin, H. Ramezani, T. Eichelkraut, T. Kottos, H. Cao, and D. N. Christodoulides, Unidirectional invisibility induced by PT-symmetric periodic structures, Phys. Rev. Lett. 106, 213901 (2011).

\bibitem{NPBGPT4} L. Feng, Y.-L. Xu, W. S. Fegadolli, M.-H. Lu, J. E. B. Oliveira, V. R. Almeida, Y.-F. Chen, and A. Scherer, Experimental demonstration of a unidirectional reflectionless parity-time metamaterial at optical frequencies, Nat. Mater. 12, 108-113 (2012).

\bibitem{KKTheo0} A. Akyurtlu and A.-G. Kussow, Relationship between the Kramers-Kronig relations and negative index of refraction, Phys. Rev. A 82, 055802 (2010).

\bibitem{KKExp1} D. Ye, C. Cao, T. Zhou, J. Huangfu, G. Zheng and L. Ran, Observation of reflectionless absorption due to spatial Kramers-Kronig profile, Nat. Commun. 8, 51 (2017).

\bibitem{KKLoh2} S. Longhi, Wave reflection in dielectric media obeying spatial Kramers-Kronig relations, EPL 112, 64001 (2015).

\bibitem{KKTheo1} T. G. Philbin, All-frequency reflectionlessness, J. Opt. 18, 01LT01 (2016).

\bibitem{KKTheo2} S. A. R. Horsley, M. Artoni, and G. C. La Rocca, Reflection of waves from slowly decaying complex permittivity profiles, Phys. Rev. A 94, 063810(2016).

\bibitem{KKTheo3} C. G. King, S. A. R. Horsley, and T. G. Philbin, Perfect transmission through disordered media, Phys. Rev. Lett. 118, 163201 (2017).

\bibitem{KKLoh3} S. Longhi, Kramers-Kronig potentials for the discrete Schr\"{o}dinger equation, Phys. Rev. A 96, 042106 (2017).

\bibitem{KKLoh4} S. Longhi, Reflectionless and invisible potentials in photonic lattices, Opt. Lett. 42, 3229-3232 (2017).

\bibitem{KKTheo4} C. G. King, S. A. R. Horsley and T. G. Philbin, Zero reflection and transmission in graded index media, J. Opt. 19, 085603 (2017).

\bibitem{kk0} L. D. Landau and E. M. Lifshitz, Electrodynamics of continuous media (Butterworth-Heinemann, 2004).

\bibitem{TMM0} M. Born and E. Wolf, Principles of optics (Cambridge University Press, Cambridge, 1980), 6th ed.

\bibitem{TMM1} M. Artoni, G. La Rocca, and F. Bassani, Resonantly absorbing one-dimensional photonic crystals, Phys. Rev. E 72, 046604 (2005).

\bibitem{TMM2} Y. Zhang, Y. Xue, G. Wang, C.-L. Cui, R. Wang, and J.-H. Wu, Steady optical spectra and light propagation dynamics in cold atomic samples with homogeneous or inhomogeneous densities, Opt. Experess 19, 2111-2019 (2011).

\bibitem{chip1} R. Folman, P. Kruger, D. Cassettari, B. Hessmo, T. Maier, and J. Schmiedmayer,
Controlling cold atoms using nanofabricated surfaces: atom chips,
Phys. Rev. Lett. 84, 4749-4752 (2000).

\bibitem{chip2} M. F. Riedel, P. Bohi, Y. Li, T. W. Hansch, A. Sinatra, and P. Treutlein,
Atom-chip-based generation of entanglement for quantum metrology,
Nature 464, 1170-1173 (2010).

\bibitem{chip3} J. D. Carter and J. D. D. Martin, Coherent manipulation of cold Rydberg atoms
near the surface of an atom chip, Phys. Rev. A 88, 043429 (2013).

\bibitem{chip4} H. Hattermann, D. Bothner, L. Y. Ley, B. Ferdinand, D. Wiedmaier, L. Sarkany,
R. Kleiner, D. Koelle, and J. Fortagh, Coupling ultracold atoms to a
superconducting coplanar waveguide resonator, Nat. Commun. 8, 2254
(2017).

\end{thebibliography}
\end{document}